\documentstyle[twoside,fleqn,espcrc2,epsfig,amssymb]{article}

\newcommand{\myvspace}{\vspace{.5ex}}

\def\rmd{{\rm d}}

\def\rme{{\rm e}}
\def\rmO{{\rm O}}

\def\defeq{\mathrel{\mathop=^{\rm def}}}
\def\proof{\noindent{\sl Proof:}\kern0.6em}

\def\frac#1#2{\hbox{$#1\over#2$}}
\def\dual{\mathstrut^*\kern-0.1em}

\def\lvec#1{\setbox0=\hbox{$#1$}
    \setbox1=\hbox{$\scriptstyle\leftarrow$}
    #1\kern-\wd0\smash{
    \raise\ht0\hbox{$\raise1pt\hbox{$\scriptstyle\leftarrow$}$}}
    \kern-\wd1\kern\wd0}
\def\rvec#1{\setbox0=\hbox{$#1$}
    \setbox1=\hbox{$\scriptstyle\rightarrow$}
    #1\kern-\wd0\smash{
    \raise\ht0\hbox{$\raise1pt\hbox{$\scriptstyle\rightarrow$}$}}
    \kern-\wd1\kern\wd0}

\def\nab#1{{\nabla_{#1}}}
\def\nabstar#1{\nabla\kern-0.5pt\smash{\raise 4.5pt\hbox{$\ast$}}
               \kern-4.5pt_{#1}}
\def\drv#1{{\partial_{#1}}}
\def\drvstar#1{\partial\kern-0.5pt\smash{\raise 4.5pt\hbox{$\ast$}}
               \kern-5.0pt_{#1}}

\def\psibar{\overline{\psi}}

\def\rhoprime{\rho\kern1pt'}
\def\rhobar{\bar{\rho}}
\def\rhobarprime{\rhobar\kern1pt'}
\def\rhobartilde{\kern2pt\tilde{\kern-2pt\rhobar}}
\def\rhobartildeprime{\kern2pt\tilde{\kern-2pt\rhobar}\kern1pt'}

\def\zetabar{\bar{\zeta}}
\def\zetaprime{\zeta\kern1pt'}
\def\zetabarprime{\zetabar\kern1pt'}
\def\zetar{\zeta_{\raise-1pt\hbox{\sixrm R}}}
\def\zetabarr{\zetabar_{\raise-1pt\hbox{\sixrm R}}}
\def\phieff{\phi_{\rm eff}}
\def\phiimpr{\phi_{\kern0.5pt\hbox{\sixrm I}}}

\def\ar{A_{\mbox{\scriptsize{\rm R}}}}
\def\vr{V_{\mbox{\scriptsize{\rm R}}}}
\def\pr{P_{\mbox{\scriptsize{\rm R}}}}

\def\dirac#1{\gamma_{#1}}
\def\diracstar#1#2{
    \setbox0=\hbox{$\gamma$}\setbox1=\hbox{$\gamma_{#1}$}
    \gamma_{#1}\kern-\wd1\kern\wd0
    \smash{\raise4.5pt\hbox{$\scriptstyle#2$}}}

\def\ba{b_{\rm A}}
\def\bv{b_{\rm V}}
\def\bp{b_{\rm P}}

\def\ca{c_{\rm A}}
\def\cv{c_{\rm V}}
\def\csw{c_{\rm sw}}

\def\fa{f_{\rm A}}
\def\ga{g^{}_{\rm A}}

\def\fIaa{f_{\rm AA}^{\rm I}}

\def\fp{f_{\rm P}}
\def\gp{g^{}_{\rm P}}

\def\fIv{f_{\rm V}^{\rm I}}
\def\f1{f_1}

\def\tr{\,\hbox{tr}\,}

\def\Sf{S_{\rm F}}
\def\Seff{S_{\rm eff}}

\def\op#1{{\cal O}_{\rm #1}}
\def\opprime#1{\setbox0=\hbox{${\cal O}$}\setbox1=\hbox{${\cal O}_{\rm #1}$}
    {\cal O}_{\rm #1}\kern-\wd1\kern\wd0
    \smash{\raise4.5pt\hbox{\kern1pt$\scriptstyle\prime$}}\kern1pt}

\def\ophatprime#1{\setbox0=\hbox{$\widehat{\cal O}$}
    \setbox1=\hbox{$\widehat{\cal O}_{\rm #1}$}
    \widehat{\cal O}_{\rm #1}\kern-\wd1\kern\wd0
    \smash{\raise4.5pt\hbox{\kern1pt$\scriptstyle\prime$}}\kern1pt}

\def\bopprime#1{\setbox0=\hbox{${\cal O}$}\setbox1=\hbox{${\cal O}_{\rm #1}$}
    {\cal L}_{\rm #1}\kern-\wd1\kern\wd0
    \smash{\raise4.5pt\hbox{\kern1pt$\scriptstyle\prime$}}\kern1pt}

\def\blagprime#1{\setbox0=\hbox{${\cal B}$}\setbox1=\hbox{${\cal B}_{#1}$}
    {\cal B}_{#1}\kern-\wd1\kern\wd0
    \smash{\raise5.2pt\hbox{\kern1pt$\scriptstyle\prime$}}\kern1pt}

\def\mq{m_{\rm q}}

\def\mr{m_{{\mbox{\scriptsize{\rm R}}}}}
\def\mc{m_{\rm c}}

\def\za{Z_{\rm A}}
\def\zp{Z_{\rm P}}
\def\zv{Z_{\rm V}}

\def\zm{Z_{\rm m}}

\def\zphi{Z_{\phi}}

\def\msbar{{\rm \overline{MS\kern-0.05em}\kern0.05em}}

\def\kc{\kappa_{\rm c}}

\title{%
\vspace{-3.1cm}
\begin{flushleft}
       {\normalsize DESY 96--157}    \\[-0.2cm]
       {\normalsize CERN--TH/96--217}  \\[-0.2cm]
       {\normalsize August 1996}   \\
\end{flushleft}
       \vspace{0.7cm}
Some new results in O($a$) improved lattice QCD
\thanks{Talks presented by S.\,Sint, R.\,Sommer and H.\,Wittig at {\sl
   Lattice\,'96}, St.\,Louis, USA, 4--8 June 1996.}}

\author{Martin L\"uscher\address{Deutsches Elektronen-Synchrotron DESY, 
                      Notkestra{\ss}e 85, D-22603 Hamburg, Germany},
        Stefan Sint\address{Max-Planck-Institut f\"ur Physik, F\"ohringer 
                      Ring 6, D-80805 M\"unchen, Germany},
        Rainer Sommer\address{CERN, Theory Division, CH-1211 Gen\`eve 23,
                      Switzerland}$^{\rm ,d}$,
        Peter Weisz$^{\rm b}$,
        Hartmut Wittig\address{DESY-IfH Zeuthen, Platanenallee 6,
                       D-15738 Zeuthen, Germany}$^{,\,}$\address{HLRZ, c/o
                       Forschungszentrum J\"ulich, 
                       D-52425 J\"ulich, Germany}
       and
        Ulli Wolff\address{Humboldt-Universit\"at, Institut f\"ur Physik,                               Invalidenstra{\ss}e 110, D-10099 Berlin, Germany}}

\begin{document}

\newcommand{\ewxy}[2]{\setlength{\epsfxsize}{#2}\epsfbox[30 30 640 640]{#1}}

\begin{abstract}
  It is shown how on-shell O($a$) improvement can be implemented
  non-perturbatively in lattice QCD with Wilson quarks.  Improvement
  conditions are obtained by requiring the PCAC relation to hold
  exactly in certain matrix elements. These are derived from the QCD
  Schr\"odinger functional which enables us to simulate directly at
  vanishing quark masses.  In the quenched approximation and for bare
  couplings in the range $0\leq g_0\leq 1$, we determine the improved
  action, the improved axial current, the additive renormalization of
  the quark mass and the isospin current normalization constants $\za$
  and $\zv$.
\end{abstract}
\maketitle


\section{INTRODUCTION}

The leading cutoff effects in lattice QCD with Wilson quarks
\cite{Wilson} are proportional to the lattice spacing $a$ and can be
rather large for typical values of the Monte Carlo (MC) simulation
parameters~\cite{letter}.  Decreasing the lattice spacing at constant
physical length scales means larger lattices and therefore rapidly
increasing costs of the MC simulations.

An alternative method to reduce cutoff effects in lattice field
theories is due to Symanzik~\cite{SymanzikI,SymanzikII}.  He has shown
that the approach of Green's functions to their continuum limit can be
accelerated by using an improved action and improved (composite)
fields.  A considerable simplification is achieved if the improved
continuum approach is only required for on-shell quantities such as
particle masses 
and matrix elements of improved fields between physical states\,[5--7].

Symanzik improvement may be viewed as an extension of the
renormalization programme to the level of irrelevant operators. While
the structure of the possible counterterms is dictated by the
symmetries, their coefficients have to be fixed by appropriate
improvement conditions. Although they can be estimated in perturbation
theory, a non-perturbative determination of the improvement
coefficients through MC simulations is clearly preferable.

To O($a$), we have recently carried out the non-perturbative
improvement programme in quenched lattice QCD, thus leaving residual
cutoff effects of O($a^2$) only.  The detailed results are given
elsewhere~[7--10].
In this report we emphasize the general concepts and give a short
account of the main results.


We first review on-shell improvement in the framework of Symanzik's
local effective theory.  The set-up in finite space-time volume with
Schr\"odinger functional boundary conditions is introduced in
sect.\,3.  We are then in the position to state the improvement
conditions and discuss their evaluation in MC simulations (sect.\,4).
Finally we present our results for the critical quark mass
(subsect.\,4.2) and the current normalization constants in the O($a$)
improved theory~(sect.\,5).


\section{ON-SHELL IMPROVEMENT}
\subsection{Lattice QCD with Wilson quarks}

In this section we consider QCD on an infinitely extended lattice with
two degenerate light Wilson quarks of bare mass $m_0$~\cite{Wilson}.
The action is the sum of the usual Wilson plaquette action and the
quark action
\begin{equation}
 \Sf[U,\psibar,\psi\,]=a^4\sum_{x}\psibar(x)(D+m_0)\psi(x),
 \label{e_quark}
\end{equation}
where $a$ denotes the lattice spacing.  The Wilson-Dirac operator
\begin{equation}
  D=\frac12\sum_{\mu=0}^3\bigl[(\nabstar\mu
  +\nab\mu)\dirac\mu-a\nabstar\mu\nab\mu\bigr],
\label{e_Wilson-Dirac}
\end{equation}
contains the lattice covariant forward and backward derivatives,
$\nab\mu$ and $\nabstar\mu$.  The last term in
eq.\,(\ref{e_Wilson-Dirac}) eliminates the unwanted doubler states but
also breaks chiral symmetry.  As a consequence, both additive and
multiplicative renormalization of the quark mass are necessary,
i.e.~any renormalized quark mass $\mr$ is of the form
\begin{equation}
 \mr=\zm \mq,\qquad \mq=m_0-\mc,
\end{equation}
where $\mc$ is the so-called critical quark mass. 

Chiral symmetry violation is more directly seen by studying the
conservation of the isovector axial current $A_\mu^a$. The current and
the associated axial density on the lattice are defined through
%
%
%
\begin{eqnarray}
  A_\mu^a(x)&=&\psibar(x)\dirac\mu\dirac5{{\tau^a}\over{2}}\psi(x),
  \label{e_axialcurrent}\\
  P^a(x)&=&\psibar(x)\dirac5{{\tau^a}\over{2}}\psi(x),
\end{eqnarray}
where $\tau^a$ are Pauli matrices
acting on the flavour index of the quark field. The PCAC relation
%
%
\begin{eqnarray}
  \tilde{\partial}_\mu A_\mu^a(x) &=& 2mP^a(x) + \rmO(a), \\
  \tilde{\partial}_\mu &=& \frac{1}{2}(\drvstar{\mu}+\drv{\mu}),
  \label{e_PCAC} 
\end{eqnarray}
then includes an error term of order~$a$, which can be rather large on
the accessible lattices\,\cite{letter}.

The isospin symmetry remains unbroken on the lattice and there exists
an associated conserved vector current. However, it is often
advantageous to use the current which is strictly local,
\begin{equation}
 V_\mu^a(x)=\psibar(x)\dirac\mu{{\tau^a}\over{2}}\psi(x).
\end{equation}
The conservation of this current is then also violated by cutoff
effects, and a finite renormalization is required to ensure that the
associated charge takes half-integral values.

	
\subsection{Symanzik's local effective theory}

Near the continuum limit the lattice theory can be described in terms
of a local effective theory~\cite{SymanzikII},
\begin{equation}
 \Seff=S_0+a S_1+a^2 S_2+\ldots,
 \label{e_Seff}
\end{equation}
where $S_0$ is the action of the continuum theory, defined e.g.~on a
lattice with spacing $\varepsilon\ll a$.  The terms $S_k$,
$k=1,2,\ldots$, are space-time integrals of lagrangians ${\cal
  L}_k(x)$.  These are given as general linear combinations of local
gauge invariant composite fields which respect the exact symmetries of
the lattice theory and have canonical dimension $4+k$.  We use the
convention that explicit (non-negative) powers of the quark mass $m$
are included in the dimension counting.  A possible basis of fields
for the lagrangian ${\cal L}_1(x)$ then reads
\begin{eqnarray}
{\cal O}_1 &=&\psibar\,\sigma_{\mu\nu}F_{\mu\nu}\psi,\nonumber\\[.2ex]
{\cal O}_2 &=&\psibar\,D_{\mu}D_{\mu}\psi
           +\psibar\,\lvec{D}_{\mu}\lvec{D}_{\mu}\psi,\nonumber\\[.2ex]
{\cal O}_3 &=& m\tr\!\left\{F_{\mu\nu}F_{\mu\nu}\right\},
              \label{e_counterterms}\\[.2ex]
{\cal O}_4 &=& m\left\{\psibar\,\dirac{\mu}D_{\mu}\psi
               -\psibar\,\lvec{D}_{\mu}\dirac{\mu}\psi\right\},
              \nonumber\\[.2ex]
{\cal O}_5 &=& m^2\psibar\psi,\nonumber
\end{eqnarray} 
where $F_{\mu\nu}$ is the field tensor and
$\sigma_{\mu\nu}=\frac{i}2[\dirac\mu,\dirac\nu]$.

When considering correlation functions of local gauge invariant fields
the action is not the only source of cutoff effects.  If $\phi(x)$
denotes such a lattice field (e.g.~the axial density or the isospin
currents of subsect.\,2.1), one expects the connected $n$-point
function
\begin{equation}
  G_n(x_1,\ldots,x_n)=(\zphi)^n
  \left\langle\phi(x_1)\ldots\phi(x_n)\right\rangle_{\rm con}
\end{equation}
to have a well-defined continuum limit, provided the renormalization
constant $\zphi$ is correctly tuned and the space-time arguments
$x_1,\ldots,x_n$ are kept at a physical distance from each other.

In the effective theory the renormalized lattice field $\zphi\phi(x)$
is represented by an effective field,
\begin{equation}
  \phieff(x)=\phi_0(x)+a\phi_1(x)+a^2\phi_2(x)+\ldots,
\end{equation} 
where the $\phi_k(x)$ are linear combinations of composite, local
fields with the appropriate dimension and symmetries.  For example, in
the case of the axial current~(\ref{e_axialcurrent}), $\phi_1$ is
given as a linear combination of the terms,
\begin{eqnarray}
  (\op{6})_{\mu}^a &=&
  \psibar\,\dirac{5}\frac{1}{2}\tau^a\sigma_{\mu\nu}
      \bigl[{D}_{\nu}-\lvec{D}_{\nu}\bigr]\psi,
  \nonumber\\[.5ex]
  (\op{7})_{\mu}^a&=&\psibar\,\frac{1}{2}\tau^a\dirac{5}
                  \bigl[{D}_{\mu}+\lvec{D}_{\mu}\bigr]\psi,
  \label{e_impr_current}\\[.5ex]
  (\op{8})_{\mu}^a&=&m\psibar\,\dirac{\mu}\dirac{5}\frac{1}{2}\tau^a\psi.
  \nonumber
\end{eqnarray}
The convergence of $G_n(x_1,\ldots,x_n)$
to its continuum limit can now be studied in the 
effective theory,
\begin{eqnarray}
  \lefteqn{G_n(x_1,\ldots,x_n) 
  =\left\langle\phi_0(x_1)\ldots\phi_0(x_n)\right\rangle_{\rm con}}
  \nonumber\\[.3ex]
  &&\mbox{}-a\int\rmd^4y\,\left\langle\phi_0(x_1)\ldots\phi_0(x_n)
  {\cal L}_1(y)\right\rangle_{\rm con}
  \nonumber\\
  &&\mbox{}+a\sum_{k=1}^n
  \left\langle\phi_0(x_1)\ldots\phi_1(x_k)\ldots\phi_0(x_n)
  \right\rangle_{\rm con}
  \nonumber\\
  &&\hphantom{0123456}+\rmO(a^2),
 \label{e_continuum_approach}
\end{eqnarray}
where the expectation values on the right hand side 
are to be taken in the continuum theory with action $S_0$.

\subsection{Using the field equations}

For most applications, it is sufficient to
compute on-shell quantities such as particle
masses, S-matrix elements and correlation functions
at space-time arguments which are separated by
a physical distance. It is then possible to make use of the
field equations to reduce first the number of 
basis fields in the effective lagrangian ${\cal L}_1$
and, in a second step, also in the O($a$) counterterm
$\phi_1$ of the effective composite fields.

If one uses the field equations in the lagrangian ${\cal L}_1$ under
the space-time integral in eq.\,(\ref{e_continuum_approach}), the
errors made are contact terms that arise when $y$ comes close to one
of the arguments $x_1,\ldots,x_n$.  Taking into account the dimensions
and symmetries, one easily verifies that these contact terms must have
the same structure as the insertions of $\phi_1$ in the last term of
eq.\,(\ref{e_continuum_approach}). Using the field equations in 
${\cal L}_1$ therefore just means a redefinition of the coefficients
in the counterterm $\phi_1$.

It turns out that one may eliminate two of the terms in
eq.\,(\ref{e_counterterms}). A possible choice is to stay with the
terms ${\cal O}_1$, ${\cal O}_3$ and ${\cal O}_5$, which yields the
effective continuum action for on-shell quantities to order $a$.
Having made this choice one may apply the field equations once again
to simplify the term $\phi_1$ in the effective field as well.  In the
example of the axial current it is then possible to eliminate the
term~$\op{6}$ in eq.\,(\ref{e_impr_current}).


\subsection{Improved lattice action and fields}

The on-shell O($a$) improved lattice action is obtained by
adding a counterterm to the unimproved lattice action
such that the action $S_1$ in the effective
theory is cancelled in on-shell amplitudes. 
This can be achieved by adding lattice representatives
of the terms ${\cal O}_1$, ${\cal O}_3$ and ${\cal O}_5$
to the unimproved lattice lagrangian, with coefficients that
are functions of the bare coupling, $g_0$, only.
Leaving the discussion of suitable improvement conditions to
sect.\,4, we here note that the fields 
${\cal O}_3$ and ${\cal O}_5$ already appear in the unimproved
theory and thus merely lead to a reparametrization of 
the bare parameters $g_0$ and $m_0$. In the following, 
we will not further consider these terms. 
For a discussion of their relevance
we refer the reader to ref.~\cite{paperI}.

We choose the standard discretization $\widehat{F}_{\mu\nu}$
of the field tensor\,\cite{paperI} and add the improvement term
to the Wilson-Dirac operator~(\ref{e_Wilson-Dirac}),
\begin{equation}
 D_{\rm impr}=D+\csw\,{{ia}\over{4}}\sigma_{\mu\nu}\widehat{F}_{\mu\nu}.
 \label{e_dimpr}
\end{equation} 
With properly chosen coefficient $\csw(g_0)$ 
this yields the on-shell O($a$) improved lattice action which 
has first been proposed by Sheikholeslami and Wohlert~\cite{SW}.

In perturbation theory, the coefficient $\csw$ 
has been computed to one-loop order by Wohlert~\cite{Wohlert}.
His result, $\csw=1+\csw^{(1)}g_0^2+\rmO(g_0^4)$, with
$ \csw^{(1)} = 0.26590(7)$, 
has recently been confirmed to three significant digits~\cite{paperII}. 

The O($a$) improved isospin currents and the axial density can
be parametrized as follows,
\begin{eqnarray}
 (\ar)_\mu^a\!\!\!&=&\!\!\!
 \za(1+\ba a\mq)\bigl\{A_\mu^a+a\ca\tilde{\partial}_\mu
       P^a\bigr\},\nonumber\\[.5ex]
 (\vr)_\mu^a\!\!\!&=&\!\!\!
 \zv(1+\bv a\mq)\bigl\{V_\mu^a+a\cv\tilde{\partial}_\nu
       T_{\mu\nu}^a\bigr\},\nonumber\\
  [.5ex]
 (\pr)^a\!\!\!&=&\!\!\!
 \zp(1+\bp a\mq)P^a\,.
\end{eqnarray}
Here all the fields on the right hand sides have been defined in
subsect.\,2.1 except the tensor density
$T^a_{\mu\nu}=i\psibar\sigma_{\mu\nu}\frac12\tau^a\psi$.  We have
included the normalization constants $Z_{\rm A,V,P}$ which have to be
fixed by appropriate normalization conditions (cf.~sect.\,5).  Again,
the improvement coefficients $b_{\rm A,V,P}$ and $c_{\rm A,V}$ are
functions of $g_0$ only. At tree level of perturbation theory we have
$\ba=\bp=\bv=1$ and $\ca=\cv=0$~\cite{HeatlieEtAl,paperII}. For $\ca$
also the one-loop result,
\begin{equation}
\ca= -0.00756(1)\times g_0^2+\rmO(g_0^4),
 \label{e_cApert}
\end{equation}
has recently been obtained~\cite{paperII}.

\subsection{The PCAC relation}

We assume for the moment that on-shell O($a$) improvement has been
fully implemented, i.e.~the improvement coefficients are assigned
their correct values. If ${\cal O}$ denotes a renormalized on-shell
O($a$) improved field localized in a region not containing $x$, we
thus expect that the PCAC relation
%
\begin{equation}
    \tilde{\partial}_\mu
    \langle(\ar)_\mu^a(x)\,{\cal O}\rangle=
     2\mr\langle(\pr)^a(x)\,{\cal O}\rangle,
 \label{e_PCAC_impr}
\end{equation}
holds up to corrections of order $a^2$. 
At this point we note that the field ${\cal O}$ need not
be improved for this statement to be true. To see this
we use again Symanzik's local effective theory and
denote the O($a$) correction term
in ${\cal O}_{\rm eff}$ by $\phi_1$.
Eq.\,(\ref{e_PCAC_impr}) then receives an order $a$ contribution
\begin{equation} 
   a\bigl\langle
  \left\{\partial_\mu(\ar)^a_{\mu}(x)-2\mr(\pr)^a(x)\right\}
  \phi_1\bigr\rangle,
 \label{e_PCACcorrection}
\end{equation}
which is to be evaluated in the continuum theory.
The PCAC relation holds exactly in this limit and
the extra term~(\ref{e_PCACcorrection}) thus
vanishes.


\section{THE SCHR\"ODINGER FUNCTIONAL}


\subsection{Definitions}

The space-time lattice is now taken to be a discretized hyper-cylinder
of length $T$ and circumference $L$.  In the spatial directions the
quantum fields are $L$-periodic, whereas in the Euclidean time
direction inhomogeneous Dirichlet boundary conditions are imposed as
follows.  The spatial components of the gauge field are required to
satisfy
\begin{equation}
  \left.U(x,k)\right|_{x_0=0}=\exp(aC_k), 
 \quad C_k=i\phi/L,
 \label{e_CCprime}
\end{equation}
with $\phi={\rm diag}(\phi_1,\phi_2,\phi_3)$, and an analogous
boundary condition with $C'$ is imposed at $x_0=T$.

With the projectors $P_\pm=\frac12(1\pm\dirac0)$, the boundary
conditions for the quark and antiquark fields read
\begin{eqnarray}
  P_{+}\psi|_{x_0=0}=\rho,
  &&
  P_{-}\psi|_{x_0=T}=\rhoprime,\\[1ex]
  \psibar P_{-}|_{x_0=0}=\rhobar,
  &&
  \psibar P_{+}|_{x_0=T}=\rhobarprime.
\end{eqnarray}
The functional integral in this situation~\cite{alphaI,StefanI},
\begin{equation} 
  {\cal Z}[C',\rhobarprime,\rhoprime;C,\bar\rho,\rho]=
  \int_{\rm fields}\,\rme^{-S},
\end{equation}
is known as the QCD Schr\"odinger functional~(SF).  Concerning the
(unimproved) action $S$, we note that its gauge field part has the
same form as in infinite volume.  The quark action is again given by
eq.\,(\ref{e_quark}), provided one formally extends the
quark and antiquark fields to Euclidean times $x_0<0$ and $x_0>T$ by
``padding" with zeros~\cite{paperI}. However, we will use a slightly
more general covariant lattice derivative,
\begin{equation}
   \nab{\mu}\psi(x)=
  {1\over a}\bigl[\lambda_{\mu}U(x,\mu)\psi(x+a\hat{\mu})-\psi(x)\bigr],
\end{equation}
where $\lambda_0=1$ and $\lambda_k=\exp(ia\theta/L)$.  This
modification of the covariant derivative is equivalent to demanding
spatial periodicity of the quark fields up to the phase
$\exp(i\theta)$.  We thus have the angle $\theta$ as an additional
parameter that plays a r\^ole in the improvement condition for the
coefficient $\ca$\,\cite{paperIII}.

We are now prepared to define the expectation values of any
product $\cal O$ of fields by
\begin{equation}
  \langle{\cal O}\rangle=
  \left\{{1\over{\cal Z}}
  \int_{\rm fields}\,{\cal O}\,
  \rme^{-S}\right\}_
  {\rhobarprime=\rhoprime=\rhobar=\rho=0}.
\end{equation}
Apart from the gauge field and the quark and anti-quark fields
integrated over, $\cal O$ may involve the ``boundary fields" at time
$x_0=0$,
\begin{equation}
   \zeta({\bf x})={\delta\over\delta\rhobar({\bf x})},
  \qquad\kern3.5ex
  \zetabar({\bf x})=-{\delta\over\delta\rho({\bf x})},
\end{equation}
and similarly the fields at $x_0=T$. Note that the functional
derivatives only act on the Boltzmann factor, because the functional
measure is independent of the boundary values of the fields.

\subsection{Continuum limit and improvement 
 of the Schr\"odinger functional}
\myvspace
Based on the work of Symanzik~\cite{SchrodingerI,SchrodingerII} and
explicit calculations to one-loop order of perturbation
theory~\cite{alphaI,StefanI} one expects that the SF is renormalized
if the coupling constant and the quark masses are renormalized in the
usual way and the quark boundary fields are scaled with a
logarithmically divergent renormalization constant.

As in the case of the infinite volume theory discussed in
subsect.\,2.2, the cutoff dependence of the SF may be described by a
local effective theory. An important difference is that the O($a$)
effective action $S_1$ now includes a few terms localized at the
space-time boundaries\,\cite{paperI}. Such terms then also appear in
the O($a$) improved lattice action. However, by an argument similar to
the one given at the end of subsect.\,2.5, it can be shown that they
only contribute at order $a^2$ to the PCAC relation and the chiral Ward
identity considered in sect.\,5. In the calculations reported below,
the inclusion of boundary counterterms is, therefore, not required. 




\section{IMPROVEMENT CONDITIONS}

A non-perturbative determination of $\csw$ and $\ca$ starts from
considering two different lattice artefacts, i.e.~combinations of
observables that are known to vanish in the continuum limit of the
theory. One then imposes the improvement conditions that these lattice
artefacts vanish for any finite value of the lattice spacing.  The
specific values of $\csw$ and $\ca$, where this is achieved then
define the $\rmO(a)$\,improved theory for that value of\,\,$a$.

Of course, there is a large freedom to choose improvement conditions
and -- on the non-perturbative level -- the resulting values of $\csw$
and $\ca$ depend on the choices made. The corresponding variation of
$\csw,\,\ca$ is of order $a$. This variation changes the effects of
order $a^2$ in physical observables computed after improvement.  In
principle this is irrelevant at the level of $\rmO(a)$ improvement.
Nevertheless, one ought to choose improvement conditions where such
terms have small coefficients. The improvement conditions derived
from the SF can be studied in perturbation theory. Such a study
provided essential criteria for our detailed choice of improvement
conditions. A further criterion is the statistical accuracy that can
be obtained in the MC computations of the lattice artefacts.

\subsection{Determination of $\csw$ and $\ca$}

Using the operator
\begin{equation}
 {\cal O}^a
 =a^6\sum_{\bf y,z}\zetabar({\bf y})\dirac 5{{\tau^a}\over{2}}\zeta({\bf z}),
 \label{e_O}
\end{equation}
we define the bare correlation functions
\begin{equation}
 \begin{array}{l}
   \mbox{$\displaystyle\fa(x_0)\,\,=\,\,-\frac{1}{3}
   \langle A_0^a(x)\,{\cal O}^a\rangle,$}
 \\[1ex]
   \mbox{$\displaystyle\fp(x_0)\,\,=\,\,-\frac{1}{3}
   \langle P^a(x)\,{\cal O}^a\rangle.$}
 \end{array}
\end{equation}
The same correlation functions, but with the boundary values $C$ and
$C^\prime$ interchanged, will be referred to as $\ga$ and $\gp$.

In terms of $\fa$ and $\fp$ the PCAC relation for the unrenormalized
improved axial current and density may be written in the form
\begin{equation}
 m= {{\tilde{\partial}_0\fa(x_0)
 +\ca a\drvstar0\drv0\fp(x_0)}\over{2\fp(x_0)}}.
 \label{e_m}
\end{equation} 
%
We take this as the definition of the bare current quark
mass~$m$. The renormalized quark mass $\mr$ appearing in
eq.\,(\ref{e_PCAC_impr}) is then given by
%
\begin{equation}
  \mr=m{{\za(1+\ba a\mq)}\over{\zp(1+\bp a\mq)}}+\rmO(a^2).
\end{equation}
At fixed bare parameters, $\mr$ and hence also the unrenormalized
mass~$m$ should be independent of the kinematical parameters such as
$T,\,L$ and $x_0$. This will be true up to corrections of order $a^2$,
provided $\csw$ and $\ca$ have been assigned their proper values. The
coefficients may, therefore, be fixed by imposing the condition that
$m$ has exactly the same numerical value for three different choices
of the kinematical parameters.


In detail we set 
$\theta=0$, $T=2L$,
$
  (\phi^{}_1,\phi^{}_2,\phi^{}_3)=\frac16(-\pi,0,\pi) $ and
$
        (\phi'_1,\phi'_2,\phi'_3)=\frac16(-5\pi,2\pi,3\pi)$.
Using the shorthand notation
$m=r_f(x_0)+\ca s_f(x_0)$  for eq.\,(\ref{e_m}), and, correspondingly,
$m=r_g(x_0)+\ca s_g(x_0)$ for the same equation with $\ga$ and $\gp$, 
our improvement conditions are   
\begin{equation} \label{e_impr}
  r_f(x_0)+\ca s_f(x_0)=r_g(x_0)+\ca s_g(x_0),
\end{equation}
at both $x_0=T/4$ and $x_0=3T/4$.

\begin{figure}[tp]
\vspace{-80pt}
\epsfig{file=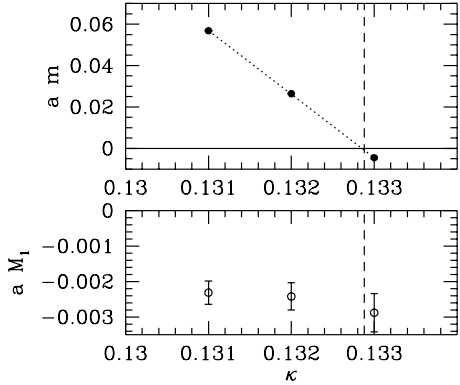,
       bbllx=68pt,%
       bblly=144pt,%
       bburx=642pt,%
       bbury=768pt,%
        clip=,%
       width=9cm}%
\vspace{-90pt}
\caption{Quark mass dependence of $M_1$ for $\beta=6.4$ and
$\csw=1.776625$.
The position of $m=0$ is marked by the dashed line. The hopping 
parameter is defined as $\kappa=1/(8+2am_0)$.
}
\label{f_m1_kappa}
\end{figure}

In order to determine $\csw$ 
we eliminate $\ca$ in eq.\,(\ref{e_impr}) and consider the combination 
\begin{equation}
 M_1 \defeq r_f(\frac34T)-r_g(\frac34T) 
  + \gamma \bigl[s_f(\frac34T)-s_g(\frac34T)\bigr],
\end{equation}
where 
\begin{equation}
\gamma=[{r_f(\frac14T)-r_g(\frac14T)}]\big/
       [{s_g(\frac14T)-s_f(\frac14T)}].
\end{equation}
The equation $M_1=0$ may then be used to determine $\csw$ for each
value of $g_0$.  In practice we suppress the $\rmO(a)$ uncertainties
in $\csw$ by a power of $g_0^2$ by equating $M_1$ to its tree level
value, $M_1^{(0)}|_{a/L}$, in the improved theory (at the finite value
of $a/L$).  We further fixed $a/L=1/8$, where both a good statistical
accuracy was achieved and the $\rmO(a)$ uncertainties of $\csw$ are
only around 2\% in a perturbative estimate.

\begin{figure}[b]
\vspace{-130pt}
\epsfig{file=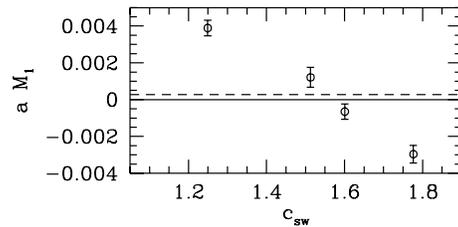,
       bbllx=68pt,%
       bblly=144pt,%
       bburx=642pt,%
       bbury=768pt,%
        clip=,%
       width=9cm}%
\vspace{-90pt}
\caption{Determination of $\csw$ at $\beta=6.4$.
The dashed line denotes the tree level value $M_1^{(0)}|_{a/L}$.
}
\label{f_m1_csw}
\end{figure}
In general we evaluated $M_1$ for vanishing mass, 
$$
m=r_f(T/2)+\gamma \, s_f(T/2). \quad 
 $$
As demonstrated in fig.\,\ref{f_m1_kappa}, finding the value of $M_1$
at $m=0$ is an easy task, since simulations of the SF are possible
for both positive and negative values of the quark
mass\,\cite{letter}.  At each of our nine values of $g_0$, this is
done for at least three values of $\csw$ and $M_1 = M_1^{(0)}|_{a/L}$
is solved for $\csw$ by linear interpolation of $M_1$ in $\csw$ (cf.
fig.\,\ref{f_m1_csw}).

In the range $0\leq g_0^2\leq 1$, the results for $\csw$ are well
represented by\,\cite{paperIII}
\begin{equation}
 \csw(g_0)={{1-0.656g_0^2-0.152g_0^4-0.054g_0^6}\over{1-0.922g_0^2}} .
 \label{e_csw_fit}
\end{equation}
\begin{figure}[tp]
\vspace{-30pt}
\epsfig{file=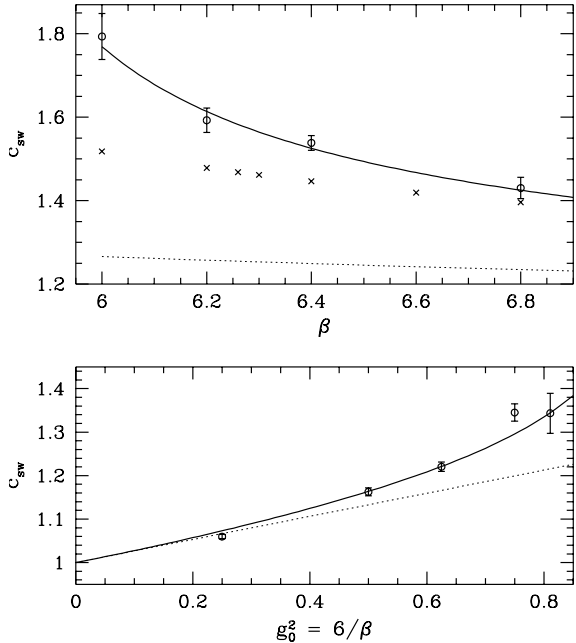,
       bbllx=68pt,%
       bblly=144pt,%
       bburx=642pt,%
       bbury=768pt,%
        clip=,%
       width=9cm}%
\vspace{-40pt}
\caption{
  Non-perturbative improvement coefficient $c_{\rm sw}$.  
  Results from one-loop bare and tadpole improved perturbation theory
  are denoted by dotted lines and crosses, respectively.}
\label{f_csw}
\end{figure}
In fig.\,\ref{f_csw} we compare our results to one-loop bare
perturbation theory and also to tadpole improved perturbation
theory\,\cite{lepenzie_92}, for which we have used
\begin{equation}
   \csw = u_0^{-3}\left[ 1 + (\csw^{(1)}-1/4)\tilde{g}^2\right],
\end{equation}
where $\tilde{g}^2=g_0^2/u_0^4$\,\cite{Parisi}. Here $u_0^4$ is the
average plaquette in infinite volume.

In a similar way\,\cite{paperIII} we obtained ($0\leq g_0^2\leq 1$)
\begin{equation}
 \ca(g_0)=-0.00756\times g_0^2{{1-0.748g_0^2}\over{1-0.977g_0^2}}.
 \label{e_ca_fit}
 \end{equation}
Note that the correction term to the axial current is rather small
and both eq.\,(\ref{e_csw_fit}) and eq.\,(\ref{e_ca_fit}) deviate 
substantially from the one-loop results except for values
of $g_0^2$ as small as $g_0^2\leq 1/2$.

\subsection{The chiral region}

As chiral symmetry is violated by lattice artefacts, there is no
precise definition of a chiral point
$\kappa=\kc\equiv1/(8+2a\mc)$ for any finite value of the lattice
spacing\,\cite{paperI}.  Rather $\kc$ has an intrinsic
uncertainty which is reduced from $\rmO(a^2)$ to $\rmO(a^3)$ by
non-perturbative improvement.

Here, we define $\kc$ as the value of $\kappa$, where $m$,
eq.\,(\ref{e_m}), vanishes for $C=C'=\theta=0, T=2L, x_0=T/2$.  To
study the $\rmO(a^3)$ effects we can then still vary the resolution
$a/L$.

Close to $\kc$ the mass $m$ is a linear function of $\kappa$ to a high
degree of accuracy. The point $\kc$ is therefore easily found by
linear interpolation (or, for $\beta \leq 6.2$, slight extrapolation).
\begin{figure}[t]
\vspace{-30pt}
\epsfig{file=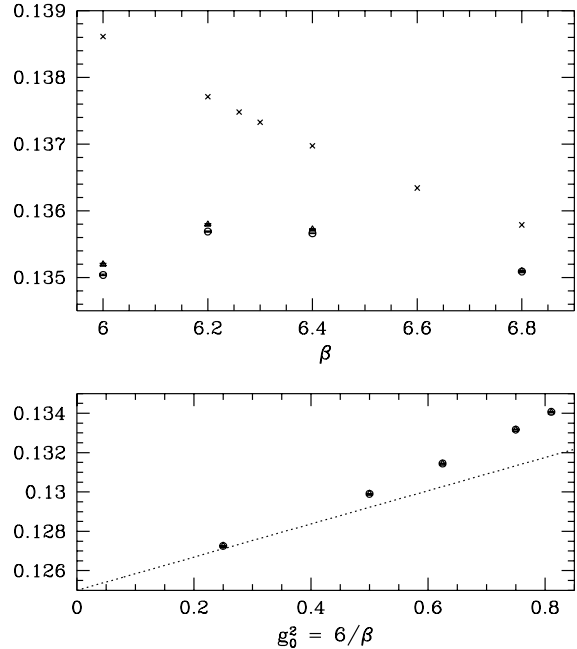,
       bbllx=68pt,%
       bblly=144pt,%
       bburx=642pt,%
       bbury=768pt,%
        clip=,%
       width=9cm}%
\vspace{-40pt}
\caption{$\kc$ determined for
  $L/a=8$ (circles), and $L/a=16$ (triangles).  The dotted line
  is the one-loop
  result\,\protect\cite{Wohlert,GabrielliEtAl,paperII}.
  Crosses represent tadpole improved perturbation theory.}
\label{f_kc}
\end{figure}
For $\beta>6.8$, the results (fig.\,\ref{f_kc}) for $a/L=1/8$ and
$a/L=1/16$ agree on the level of their statistical accuracy which is
better than $10^{-5}$. For lower $\beta$ small but significant
$\rmO(a^3)$ effects are seen.
 
Note that the pion mass (in infinite volume) will not vanish exactly
at $\kappa=\kc$ since there is no exact chiral symmetry for finite
values of $a$. However, for $\kappa=\kc$, as defined here, the pion
mass is expected to vanish up to a small lattice artefact of order
$\rmO(a^2)$.


\section{CURRENT NORMALIZATION}
\subsection{Chiral Ward identities}

For zero quark masses chiral symmetry is expected to become exact in
the continuum limit.  It has therefore been proposed to fix the
normalization constants $\za$ and $\zv$ of the isospin currents by
imposing the continuum chiral Ward identities also at finite values of
the cutoff~[19--21].

In the case of the axial current the relevant Ward identity can be
written in the form
\begin{equation}
 \int_{\partial R}\!\!\!\!\rmd\sigma_\mu(x)\epsilon^{abc}
 \langle A_\mu^a(x)A_\nu^b(y) {\cal Q}^c\rangle 
  \!=\! 2i\langle V_\nu^c(y){\cal Q}^c\rangle
\label{e_Ward}
\end{equation}
where the integral is taken over the boundary of the space-time region
$R$ containing the point $y$ and ${\cal Q}^a$ is a source located
outside $R$.  In view of on-shell improvement it is important to note
that all space-time arguments in eq.\,(\ref{e_Ward}) are at non-zero
distance from one another.

For the source field in eq.\,(\ref{e_Ward}) we choose
${\cal Q}^a=\epsilon^{abc}{\cal O}'^{b}{\cal O}^{c}$
with ${\cal O}^a$ as given in eq.\,({\ref{e_O}), 
and ${\cal O}'{}^a$ defined similarly using the primed fields.

We define the following correlation functions, using the
unrenormalized improved currents
$(A_{\rm I})_\mu^a=A_\mu^a+a\ca\tilde{\partial}_\mu P^a$ and 
$(V_{\rm I})_\mu^a=V_\mu^a+a\cv\tilde{\partial}_\nu T_{\mu\nu}$,
\begin{eqnarray}
  \fIaa(x_0,y_0)
         &=&-{{a^6}\over{6L^6}}\sum_{\bf x,y}\epsilon^{abc}\epsilon^{cde}
                    \nonumber\\
               & & \hspace{-0.3cm}\times\langle{\cal O}'^{d}
        (A_{\rm I})_0^a(x)(A_{\rm I})_0^b(y){\cal O}^{e}\rangle,
\end{eqnarray}
\vspace{-0.3cm}
\begin{equation}
  \fIv(x_0) = {{a^3}\over{6L^6}}\sum_{\bf x} i\epsilon^{abc}
                    \langle{\cal O}'^{a}(V_{\rm I})_0^b(x) 
                   {\cal O}^c\rangle,
\end{equation}
\vspace{-0.3cm}
\begin{equation}
  \f1       = -{{1}\over{3L^6}}\langle{\cal O}'^{a}{\cal O}^{a}\rangle.
\end{equation}

At $\mq=0$, and using the correct values of $\csw$ and $\ca$, 
a lattice version of the chiral Ward identity (\ref{e_Ward}) is
\begin{equation} \label{eq:za2}
 \za^2\fIaa(x_0,y_0)=\f1 +\rmO(a^2),\qquad x_0 > y_0 \, .
 \label{e_za}
\end{equation}
%
Compared to eq.\,(\ref{e_Ward}) we have set $\nu=0$ and included an
additional summation over ${\bf y}$, thus obtaining the isospin
charge. In deriving eq.\,(\ref{e_za}) we have used the fact that the
action of the latter on the chosen matrix elements can be evaluated
due to the exact isospin symmetry on the lattice.

Since the isospin symmetry remains unbroken for non-zero quark mass,
one need not restrict the normalization condition of the vector
current to the case $\mq=0$. Using similar arguments as in the
derivation of eq.\,(\ref{eq:za2}) one obtains
\begin{equation} \label{eq:zv}
 \zv(1+\bv a\mq)\fIv(x_0)=\f1 +\rmO(a^2).
\end{equation}
%
Note that the improvement coefficient $\cv$ is not needed here,
because the tensor density
does not contribute to the isospin charge.

\subsection{Numerical evaluation}

Starting from eqs.\,(\ref{eq:za2},\,\ref{eq:zv}) we impose the
following normalization conditions on the axial and vector currents
at vanishing quark mass, $\mq=0$,
\begin{eqnarray}
 \za^2\fIaa(\frac23T,\frac13T) & = & \f1, \\
 \zv\fIv(\frac12T) & = & \f1,
\end{eqnarray}
where we have set $C=C'=\theta=0$.  It is clear that any other
choice of the parameters will lead to the same result up to effects of
the order $a^2$. We will use these effects as well as those from
varying $L,\,T$ and the insertion points $x_0,\,y_0$ as an estimate of
systematic errors on the normalization constants.

Statistical errors on $\za$ and $\zv$ were found to be much smaller
than those on $\fIaa,\,\fIv$ and $f_1$, due to the strong cancellation
of correlations in the ratios $f_1/\fIaa$ and $f_1/\fIv$. A detailed
analysis of systematic errors at $\beta=6.4$ revealed that the spread
of results obtained from varying $C,\,C^\prime,\,\theta,\,L,\,T$ and
the insertion points was mostly well within the statistical
errors. The total error was estimated to be at the 1\% level for $\zv$
and $\za$.

Fig.\,\ref{f_zvza} shows the measured data for $\zv$ and $\za$
obtained for $L/a=8$, $T/a=18$.  They are compared to one-loop
perturbation theory, for which we used the results of
ref.\,\cite{GabrielliEtAl} which have been confirmed by one of us. The
figure shows that contact with one-loop lattice perturbation theory
can be established for $g_0^2\lesssim 0.5$. At larger couplings there
is also very good agreement between our data and one-loop tadpole
improved
perturbation theory.

\begin{figure}[t]       
\vspace{-4.cm}
\ewxy{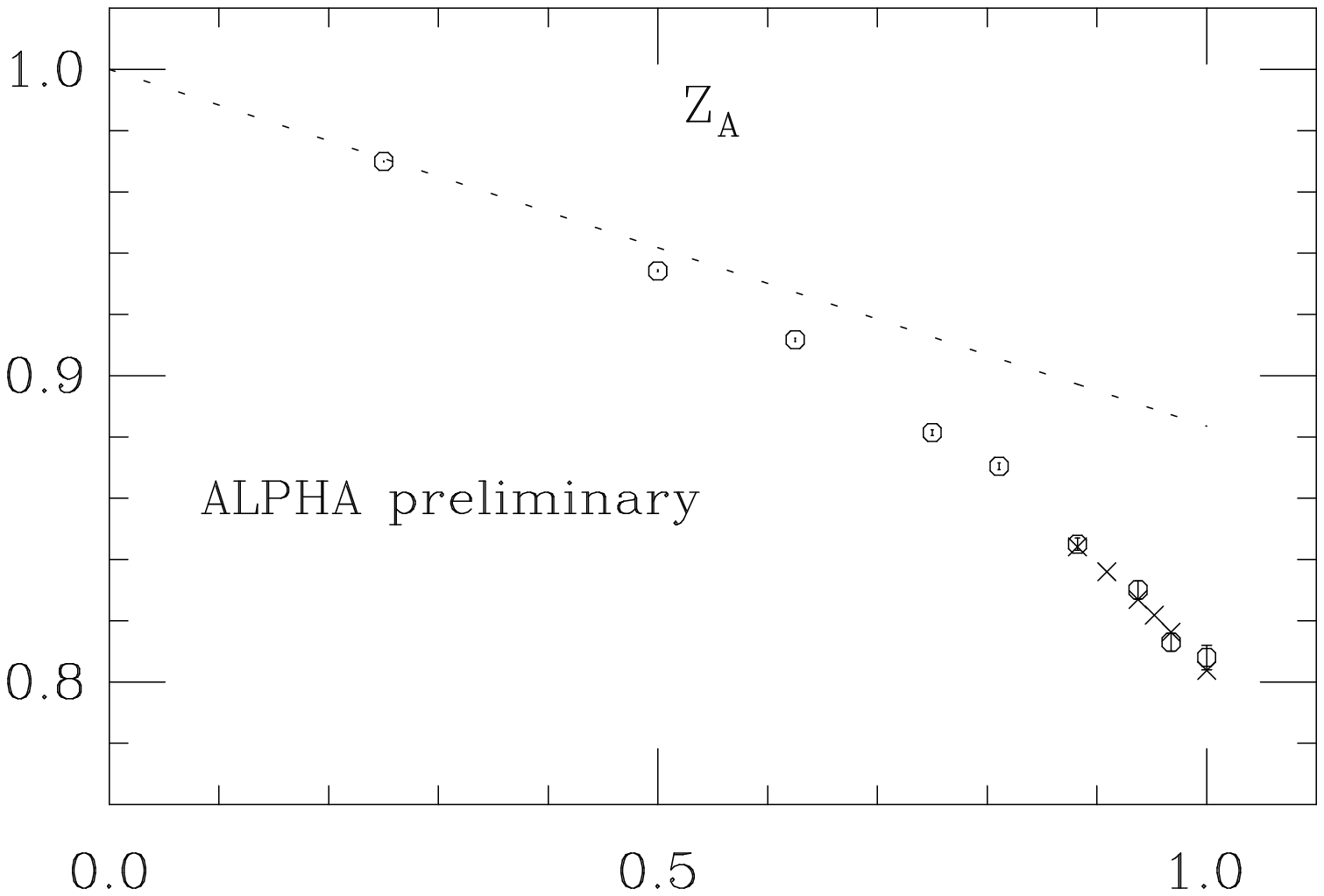}{90mm}
\vspace{-5.cm}
\ewxy{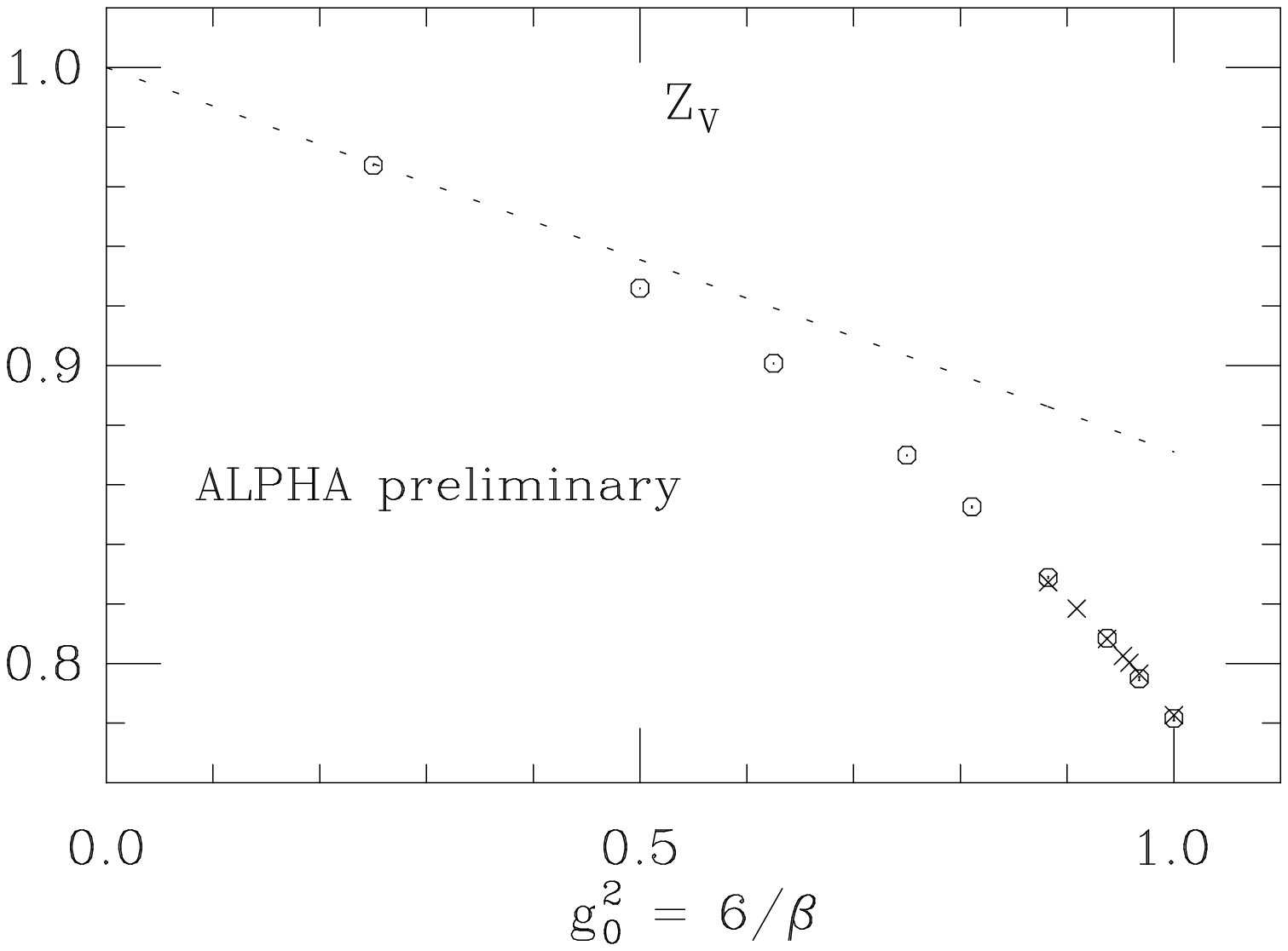}{90mm}
\vspace{-1.5cm}
\caption{$Z_A$ and $Z_V$ plotted against $g_0^2$.
Results from one-loop bare and tadpole improved perturbation
theory are denoted by dotted lines and crosses, respectively.
}
\label{f_zvza}
\end{figure}

In the case of the vector current one may extend the normalization
condition to non-vanishing quark masses, such that the ratio
$\fIv(T/2)/f_1$ yields the combination $\zv(1+\bv a\mq)$, from which
one can extract the improvement coefficient $\bv$. 
A preliminary analysis shows that $\bv$ is about 40\% above its
tree-level value of $\bv=1$ for $6.0\leq\beta\leq6.4$.
 
This procedure can, however, not be applied to the axial current,
since the mass dependence of its normalization is obscured by the
physical mass dependence arising from the axial density in the Ward
identity.


\section{CONCLUSIONS}

We have been able to implement on-shell O($a$) improvement
non-perturbatively in (quenched) lattice QCD. The improvement
coefficients $\csw$, $\ca$, the critical mass $\mc$ and the current
normalization constants $\za$ and $\zv$ have been determined for bare
couplings in the range $0\leq g_0\leq1$.  In all cases contact with
bare perturbation theory could be made at couplings around $g_0^2
\approx 0.25 - 0.5$.  The convergence of perturbation theory appears
to be speeded up significantly by tadpole improvement as shown by the
crosses in figures~\ref{f_csw}--\ref{f_zvza}. However, for values of
$\beta \le 6.8$, which is the range most relevant for present MC
computations, the quality of tadpole improved perturbation theory is
rather non-universal. While excellent agreement between perturbation
theory and our results for $\za,\zv$ is found, substantial deviations
are observed in the case of $\csw,\,\ca$ and $\kc$.

\vskip.5ex
This work is part of the ALPHA collaboration research programme. We
thank DESY for allocating computer time on the APE-Quadrics computers
for this project and are grateful to the staff of the computer center
at DESY-IfH, Zeuthen, where all numerical simulations have been
performed.


\end{document}